\documentclass[12pt]{article}

\usepackage{amssymb}
\usepackage{amsmath}

\usepackage{epsfig}

\begin{document} %


\title{
\bf Generalization of the Randall-Sundrum solution}

\author{
A.V. Kisselev\thanks{Electronic address:
alexandre.kisselev@ihep.ru} \\
{\small Institute for High Energy Physics, NRC ``Kurchatov
Institute''\!,} \\
{\small 142281 Protvino, Russian Federation}
}

\date{}

\maketitle


\thispagestyle{empty}

\begin{abstract}
The generalization of the Randall-Sundrum solution for the warp
factor $\exp[\sigma(y)]$ in the scenario with one extra coordinate
$y$, non-factorizable space-time geometry and two branes is
obtained. It is shown that the function obtained $\sigma(y)$ is
symmetric with respect to an interchange of two branes. It also
obeys the orbifold symmetry $y \rightarrow - y$ and explicitly
reproduces jumps of its derivative on both branes. This solution is
defined by the Einstein-Hilbert's equations up to a constant $C$. It
is demonstrated that different values of $C$ results in theories
with quite different spectra of the Kaluza-Klein gravitons.
\end{abstract}





\section{Introduction}

The 5-dimensional space-time with non-factorizable geometry and two
branes was suggested by Randall and Sundrum  (RS1 model)
\cite{Randall:99} as an alternative to the ADD model with flat extra
dimensions     \cite{Arkani-Hamed:98}-\cite{Arkani-Hamed:99}. Its
phenomenological implications were explored soon
\cite{Davoudiasl:00}. The model predicts an existence of heavy
Kaluza-Klein excitations (KK gravitons). These massive resonances
are intensively searched for by the LHC collaborations (see, for
instance, \cite{ATLAS:gravitons}, \cite{CMS:gravitons}).

The RS scenario is described by the following background warped
metric
\begin{equation}\label{RS_background_metric}
\quad ds^2 = e^{-2 \sigma (y)} \, \eta_{\mu \nu} \, dx^{\mu} \,
dx^{\nu} - dy^2 \;,
\end{equation}
where $\eta_{\mu\nu}$ is the Minkowski tensor with the signature
$(+,-,-,-)$, and $y$ is an extra coordinate. It is a model of
gravity in the AdS$_5$ space-time compactified to the orbifold
$S^1\!/Z_2$. There are two branes located at the fixed points of the
orbifold. The function $\sigma(y)$ in the warp factor $\exp[-2
\sigma(y)]$ was obtained to be \cite{Randall:99}
\begin{equation}\label{sigma_RS1}
\sigma_{\mathrm{RS}} (y) = \kappa |y| \;,
\end{equation}
where $\kappa$ is a parameter with a dimension of mass.

This expression is consistent with the orbifold symmetry $y
\rightarrow - y$. However, it is not symmetric with respect to the
branes. The jump of the derivative $\sigma'(y)$ on the brane $y=\pi
r_c$ does not follow from expression \eqref{sigma_RS1}
\emph{directly}, but only after taking into account periodicity
condition.%
\footnote{Here and in what follows, the \emph{prime} denotes the
derivative with respect to variable $y$.}
Moreover, a constant can be safely added to $\sigma (y)$. Thus, a
generalization of the RS solution \eqref{sigma_RS1} is needed.

In the present paper we will derive such a general solution
$\sigma(y)$ of the Einstein-Hilbert's equations which has the
following properties: (i) it has the orbifold symmetry $y
\rightarrow - y$; (ii) jumps of $\sigma'(y)$ are explicitly
reproduced on both branes; (iii) it is symmetric with respect to the
interchange of the branes; (iv) it includes a constant term.

Previously, the solution for $\sigma(y)$ was studied in
ref.~\cite{Kisselev:13}. In the present paper we reconsider and
strengthen arguments used in deriving this solution, as well as
correct expressions for $\sigma'(y)$ and 5-dimensional cosmological
constant $\Lambda$ presented in \cite{Kisselev:13}. Moreover, the
solution in \cite{Kisselev:13} was {incomplete}, since it did not
contain an additional dimensionless quantity $C$ ($0  \leqslant C
\leqslant |\kappa| \pi r_c$). As it is shown in the present paper, a
physical content of a theory depends crucially on a particular value
of $C$.

In Section~\ref{sec:2} a generalization of the Randall-Sundrum
solution \eqref{sigma_RS1} is derived, and in Section~\ref{sec:3}
properties of a new solution are discussed in detail.


\section{RS solution and its generalization}
\label{sec:2}

The classical action of the Randall-Sundrum scenario
\cite{Randall:99} is given by
\begin{align}\label{action}
S &= \int \!\! d^4x \!\! \int_{-\pi r_c}^{\pi r_c} \!\! dy \,
\sqrt{G} \, (2 \bar{M}_5^3 \mathcal{R}
- \Lambda) \nonumber \\
&+ \int \!\! d^4x \sqrt{|g^{(1)}|} \, (\mathcal{L}_1 - \Lambda_1) +
\int \!\! d^4x \sqrt{|g^{(2)}|} \, (\mathcal{L}_2 - \Lambda_2) \;,
\end{align}
where $G_{MN}(x,y)$ is the 5-dimensional metric, with $M,N =
0,1,2,3,4$, $\mu = 0,1,2,3$, and $y$ is the 5-th dimension
coordinate of the size $\pi r_c$. The quantities
\begin{equation}
g^{(1)}_{\mu\nu}(x) = G_{\mu\nu}(x, y=0) \;, \quad
g^{(2)}_{\mu\nu}(x) = G_{\mu\nu}(x, y=\pi r_c)
\end{equation}
are induced metrics on the branes, $\mathcal{L}_1$ and
$\mathcal{L}_2$ are brane Lagrangians, $G = \det(G_{MN})$, $g^{(i)}
= \det(g^{(i)}_{\mu\nu})$.

The periodicity condition, $y = y \pm 2\pi r_c$, is imposed and the
points $(x_{\mu}, y)$ and $(x_{\mu}, -y)$ are identified. So, one
gets the orbifold $S^1/Z_2$. We consider the case with two 3-branes
located at the fixed points $y = 0$ (Plank brane) and $y = \pi r_c$
(TeV brane). The SM fields are constrained to the TeV (physical)
brane, while the gravity propagates in all spatial dimensions.

From action \eqref{action} 5-dimensional Einstein-Hilbert's
equations follow
\begin{align}\label{H-E_equation}
\sqrt{|G|} & \left( \mathcal{R}_{MN} - \frac{1}{2} \, G_{MN}
\mathcal{R} \right) = - \frac{1}{4 \bar{M}_5^3} \Big[ \sqrt{|G|} \,
G_{MN}  \Lambda
\nonumber \\
&+  \sqrt{|g^{(1)}|} \, g^{(1)}_{\mu\nu} \, \delta_M^\mu \,
\delta_N^\nu \, \delta(y) \, \Lambda_1 +  \sqrt{|g^{(2)}|} \,
g^{(2)}_{\mu\nu} \, \delta_M^\mu \, \delta_N^\nu \, \delta(y - \pi
r_c) \, \Lambda_2 \Big] \;.
\end{align}
In what follows, the reduced scales will be used:
$\bar{M}_{\mathrm{Pl}} = M_{\mathrm{Pl}} /\sqrt{8\pi} \simeq
2.4\cdot 10^{18} \ \mathrm{GeV}$, and $\bar{M}_5 = M_5 /(2\pi)^{1/3}
\simeq 0.54 \, \bar{M}_5$.

In order to solve Einstein-Hilbert's equations, it is assumed that
the background metric respects 4-dimensional Poincare invariance
\eqref{RS_background_metric}. After orbifolding, the coordinate of
the extra compact dimension varies within the limits $0 \leqslant  y
\leqslant \pi r_c$. Then the 5-dimensional background metric tensor looks like%
\footnote{We ignore the backreaction of the brane term on the
space-time geometry.}
\begin{equation}\label{cov_metric_tensor}
G_{M\!N} = \left(
  \begin{array}{cc}
  g_{\mu\nu} & 0 \\
    0 & -1 \\
  \end{array}
\right) \;,
\end{equation}
where $g_{\mu\nu} = \exp(-2 \sigma) \, \eta_{\mu\nu}$. For the
background metric, the Einstein-Hilbert's equations are reduced to
the following set of two equations
\begin{align}
6 \sigma'^2 (y) &= - \frac{\Lambda}{4 \bar{M}_5^3} \;,
\label{sigma_deriv_eq} \\
3\sigma''(y) &= \frac{1}{4 \bar{M}_5^3} \, [\Lambda_1 \, \delta(y) +
\Lambda_2 \, \delta(\pi r_c - y)] \;.
\label{sigma_2nd_deriv_eq}
\end{align}
Let us note that the function $\sigma(y)$ is defined by this set of
equations up to a constant.

In between the branes (i.e. for $0 < y < \pi r_c$) we get from
\eqref{sigma_2nd_deriv_eq} that $\sigma''(y) = 0$, that results in
$\sigma'(y) = \kappa$, where $\kappa$ is a scale with a dimension of
mass.

Let us define dimensionless quantities $\lambda$, $\lambda_1$ and
$\lambda_2$ ($\lambda > 0, \lambda_{1,2} \neq 0$),
\begin{equation}\label{small_lambdas}
\Lambda = -24 \bar{M}_5^3 \kappa^2 \! \lambda  \;, \quad
\Lambda_{1,2} = 12\bar{M}_5^3 \kappa \lambda_{1,2} \;.
\end{equation}
Then we obtain
\begin{align}
\sigma'^2 (y) &= \kappa^2 \lambda \;, \label{sigma_mod_eq_1} \\
\sigma''(y) &= \kappa [\lambda_1 \, \delta(y) + \lambda_2 \ \delta(y
- \pi r_c)] \;.
\label{sigma_mod_eq_2}
\end{align}
The quantity $\kappa$ defines a magnitude of the 5-dimensional
scalar curvature.

The branes must be treated on an \emph{equal} footing. It means that
the function $\sigma(y)$ should be symmetric with respect to the
simultaneous replacements $|y| \rightleftarrows |y - \pi r_c|$,
$\lambda_1 \rightleftarrows \lambda_2$. For the interval $0
\leqslant y \leqslant \pi r_c$, the
solution of eq. \eqref{sigma_mod_eq_2} looks like%
\footnote{We omitted a term linear in $y$, since it explicitly
violets the orbifold symmetry.}
\begin{equation}\label{sigma_lambdas}
\sigma (y) = \frac{\kappa}{4} [ ( \lambda_1 - \lambda_2) (|y| - |y -
\pi r_c | ) +  ( \lambda_1 + \lambda_2) (|y| + |y - \pi r_c | ) ] +
\mathrm{\ constant} \;,
\end{equation}
where
\begin{equation}\label{lambda_relation_1}
\lambda_1 - \lambda_2 = 2 \;.
\end{equation}
Note that eq.~\eqref{lambda_relation_1} guarantees that $\sigma'(y)
= \kappa$ for $0 < y < \pi r_c$.

There are two possibilities:
\begin{itemize}
  \item brane tensions have the same sign  \\
The function $\sigma(y)$ should be symmetric with respect to the
replacement $|y| \rightarrow |y - \pi r_c|$, since under such a
replacement the branes are interchanged (the fixed point $y=0$
becomes the fixed point $y=\pi r_c$, and vice versa). Then one has
to put $\lambda_1 - \lambda_2 = 0$ that contradicts
eq.~\eqref{lambda_relation_1}. Thus, this case cannot be realized.

  \item brane tensions have the opposite signs \\
The warp function $\sigma(y)$ must be symmetric under the
simultaneous substitutions $|y| \rightarrow |y- \pi r_c|$, $\kappa
\rightarrow - \kappa$. Thus, one has to take
\begin{equation}\label{lambda_relation_2}
\lambda_1 + \lambda_2 = 0 \;.
\end{equation}
\end{itemize}

It follows from \eqref{lambda_relation_1}, \eqref{lambda_relation_2}
that the brane tensions are
\begin{equation}\label{brane_lambdas_solution}
\lambda_1 = - \lambda_2 = 1 \;.
\end{equation}
As a result, we come to the unique solution:
\begin{equation}\label{sigma_solution}
\sigma (y) = \frac{\kappa}{2} ( |y| - |y - \pi r_c | ) +
\frac{|\kappa| \pi r_c}{2} - C \;.
\end{equation}
The constant terms in \eqref{sigma_solution} are chosen in such a
way that one has
\begin{equation}\label{sigma_internal}
\sigma (y) = \kappa y - C
\end{equation}
for $\kappa>0$ within the interval $0 < y < \pi r_c$.%
\footnote{The absolute value of $\kappa$ in the second term in
\eqref{sigma_solution} is needed to ensure the symmetry with respect
to the branes, see our comments after
eq.~\eqref{sum_two_solutions}.}
Taking into account the periodicity condition and orbifold symmetry,
we put
\begin{equation}\label{C_limits}
0 \leqslant C \leqslant |\kappa| \pi r_c \;.
\end{equation}

It follows from Einstein-Hilbert's eq.~\eqref{sigma_mod_eq_2}, as
well as from \eqref{sigma_solution}, that
\begin{equation}\label{sigma_deriv_solution}
\sigma'(y) = \frac{\kappa}{2} \, [ \varepsilon(y) - \varepsilon(y -
\pi r_c) ] \;.
\end{equation}
Let us stress that the domain of definition of the function
$\varepsilon(x)$ in \eqref{sigma_deriv_solution} \emph{must be
constrained} to the region $0 < |x| \leqslant \pi r_c$. Outside this
region, one has to use the periodicity condition first in order to
define $\sigma'(y)$ correctly.%
\footnote{As one has to do with expression \eqref{sigma_RS1} to get
a correct result (for details, see Section~\ref{sec:3}).}
In particular, it means that for $0 < y_0 < \pi r_c$
\begin{equation}\label{epsilon}
\varepsilon(-y_0 - \pi r_c) = \varepsilon(- y_0 - \pi r_c +
\boldsymbol{2\pi r_c}) = 1 \;.
\end{equation}
Then we find from \eqref{sigma_deriv_solution}, \eqref{epsilon} that
$\sigma'(-y) = -\sigma' (y)$, as it should be for the derivative of
the symmetric function $\sigma(y)$, while eq.~\eqref{sigma_mod_eq_1}
says that
\begin{equation}\label{lambda_solution}
\lambda = 1 \;.
\end{equation}

In initial notations,
\begin{align}
\Lambda &= -24 \bar{M}_5^3\kappa^2 \;, \label{Lambda_fine_tuning}
\\
\Lambda_1 &= - \Lambda_2 = 12 \bar{M}_5^3 \kappa  \;.
\label{Lambdas_fine_tuning}
\end{align}
The RS1 fine tuning relations look slightly different
\cite{Randall:99},
\begin{align}
\Lambda_{\mathrm{RS}} &= -24 \bar{M}_5^3\kappa^2 \;,
\label{RS1_Lambda_fine_tuning}
\\
(\Lambda_1)_{\mathrm{RS}} &= - (\Lambda_2)_{\mathrm{RS}} = 24
\bar{M}_5^3 \kappa \;. \label{RS1_Lambdas_fine_tuning}
\end{align}
It is necessary to stress that the bulk cosmological term $\Lambda$
is given by eq.~\eqref{Lambda_fine_tuning} in between the branes ($0
< y < \pi r_c$), but it \emph{is not} defined on the branes
themselves (i.e. at $y=0, \, \pi r_c$), as it follows from
eqs.~\eqref{sigma_deriv_eq}, \eqref{sigma_deriv_solution}.%
\footnote{Since $\sigma' (y)$ is not defined for $y=n \pi r_c$, $n =
0, \pm 1, \ldots$.}
No comments were made in \cite{Randall:99} on discontinuity of
$\Lambda_{\mathrm{RS}}$ \eqref{RS1_Lambda_fine_tuning} on the
branes.

As for the brane tensions \eqref{Lambdas_fine_tuning}, they are a
factor of 2 different than that of RS1
\eqref{RS1_Lambdas_fine_tuning}. It is a consequence of the symmetry
of $\sigma(y)$ with respect to the brane points, which is absent in
the analytical solution \eqref{sigma_RS1}.

If we start from the fixed point $y = \pi r_c$ instead of the point
$y=0$, we come to the \emph{equivalent} solution related to the TeV
brane (for a while, we assume that $\kappa>0$)
\begin{equation}\label{sigma_RS_pi}
\sigma_\pi(y) = -\kappa |y - \pi r_c| + \kappa \pi r_c \;.
\end{equation}
Note that \eqref{sigma_RS_pi} and \eqref{sigma_RS1} coincide at $0 <
y < \pi r_c$. Our final formula \eqref{sigma_solution} is in fact a
half-sum of these two solutions (up to the quantity $-C$),
\begin{equation}\label{sum_two_solutions}
\sigma(y) = \frac{1}{2}[\sigma_0(y) + \sigma_\pi(y)] - C \;,
\end{equation}
where $\sigma_0(y) = \sigma_{\mathrm{RS}}(y)$ is the solution
related to the Planck brane.

One can verify that our solution $\sigma(y)$ \eqref{sigma_solution}
obeys $Z_2$ symmetry if he takes into account the periodicity in
variable $y$ (for details, see Section~\ref{sec:3}).

The expression \eqref{sigma_solution} is also symmetric with respect
to the branes. Indeed, under the replacement $y \rightarrow \pi r_c
- y$, the positions of the branes are interchanged (the point $y=0$
becomes the point $y=\pi r_c$, and vice versa), while under the
replacement $\kappa \rightarrow - \kappa$, the tensions of the
branes \eqref{Lambdas_fine_tuning} are interchanged.

Our solution \eqref{sigma_solution} can be rewritten in the form
explicitly symmetric with respect to the brane
\begin{equation}\label{sigma_solution_gen}
\sigma (y) = \frac{\kappa}{2} ( \lambda_1 |y| + \lambda_2 |y - \pi
r_c | ) + \frac{|\kappa| \pi r_c}{2} - C \;.
\end{equation}
Here $\lambda_1=1$ and $\lambda_2=-1$ are the reduced tensions of
the branes located at the points $y=0$ and $y=\pi r_c$,
respectively. Correspondingly,
\begin{equation}\label{sigma_deriv_solution_gen}
\sigma'(y) = \frac{\kappa}{2} \, [ \lambda_1 \varepsilon(y) +
\lambda_2 \varepsilon(y - \pi r_c) ] \;.
\end{equation}

Let us stress that not only the brane warp factors, but hierarchy
relations and graviton mass spectra depend drastically on a
particular value of the constant $C$ in \eqref{sigma_solution}.
Correspondingly, the parameters of the model, $\bar{M}_5$ and
$\kappa$, can differ significantly for different $C$.

From now on, it will be assumed that $\kappa > 0$, and $\pi\!\kappa
r_c \gg 1$. The hierarchy relation is given by the formula
\begin{equation}\label{hierarchy_relation}
\bar{M}_{\mathrm{Pl}}^2  = \frac{\bar{M}_5^3}{\kappa} e^{2C} \left(
1 - e^{-2\pi \kappa r_c} \right) \simeq \frac{\bar{M}_5^3}{\kappa}
\, e^{2C} \;.
\end{equation}
The interactions of the gravitons $h_{\mu\nu}^{(n)}$ with the SM
fields on the physical brane (brane 2) are given by the effective
Lagrangian
\begin{equation}\label{Lagrangian}
\mathcal{L}_{\mathrm{int}} = - \frac{1}{\bar{M}_{\mathrm{Pl}}} \,
h_{\mu\nu}^{(0)}(x) \, T_{\alpha\beta}(x) \, \eta^{\mu\alpha}
\eta^{\nu\beta} - \frac{1}{\Lambda_\pi} \sum_{n=1}^{\infty}
h_{\mu\nu}^{(n)}(x) \, T_{\alpha\beta}(x) \, \eta^{\mu\alpha}
\eta^{\nu\beta} \;,
\end{equation}
were $T^{\mu \nu}(x)$ is the energy-momentum tensor of the SM
fields, and the coupling constant of the massive modes is
\begin{equation}\label{Lambda_pi}
\Lambda_\pi \simeq \frac{\bar{M}_{\mathrm{Pl}}}{\sqrt{\exp(2\kappa
\pi r_c) - 1}}\simeq \bar{M}_{\mathrm{Pl}} \, e^{-\kappa \pi r_c}
\;.
\end{equation}

The graviton masses $m_n$ ($n = 1, 2, \ldots $) are defined from the
equation
\begin{equation}\label{masses_eq}
J_1 (a_{1n}) Y_1(a_{2n}) - Y_1 (a_{1n}) J_1(a_{2n}) = 0 \;,
\end{equation}
where
\begin{equation}\label{a_i}
a_1 = \frac{m_n}{\kappa} \, e^{-C} =
\frac{m_n}{\bar{M}_{\mathrm{Pl}}} \left( \frac{\bar{M}_5}{\kappa}
\right)^{3/2} \!, \quad a_2 = \frac{m_n}{\kappa} \, e^{\kappa \pi
r_c - C} = \frac{m_n}{\bar{M}_{\mathrm{Pl}}} \left(
\frac{\bar{M}_5}{\kappa} \right)^{3/2} \!\! e^{\kappa \pi r_c} \;.
\end{equation}
As a result, for all $m_n \ll  \bar{M}_{\mathrm{Pl}}
(\kappa/\bar{M}_5)^{3/2}$, we get
\begin{equation}\label{m_n}
m_n = x_n \bar{M}_{\mathrm{Pl}} \! \left( \frac{\kappa}{\bar{M}_5}
\right)^{3/2} \!\! e^{-\kappa \pi r_c} = x_n \Lambda_\pi \! \left(
\frac{\kappa}{\bar{M}_5} \right)^{3/2} ,
\end{equation}
where $x_n$ are zeros of the Bessel function $J_1(x)$.

By taking different values of $C$ in eq.~\eqref{sigma_solution}, we
come to quite diverse \emph{physical scenarios}. One of them ($C =
0$) is in fact the RS1 model \cite{Randall:99}. Another scheme ($C =
\kappa \pi r_c$) describes a geometry with a small curvature of
five-dimensional space-time \cite{Giudice:05}-\cite{Kisselev:06}
(RSSC model). It predicts a spectrum of the KK gravitons similar to
a spectrum of the ADD model
\cite{Arkani-Hamed:98}-\cite{Arkani-Hamed:99}. For the LHC
phenomenology of the RSSC model, see, for instance,
\cite{Kisselev:diphotons}, \cite{Kisselev:dimuons}. The scheme with
$C = \kappa \pi r_c/2$, and $\sigma (0) = - \sigma (\pi r_c) = -
\kappa \pi r_c/2$ also lead to an interesting phenomenology quite
different from that of the RS1 model. The details is a subject of a
separate publication.

Both the mass spectrum of the KK gravitons \eqref{m_n} and theirs
interaction with the SM fields \eqref{Lambda_pi} depend on $C$,
although implicitly. The point is that the parameters $M_5$ and
$\kappa$ in equations \eqref{m_n} and \eqref{Lambda_pi} \emph{do
depend} on $C$ via the hierarchy relation
\eqref{hierarchy_relation}. Indeed, the RS1 hierarchy relation ($C =
0$) looks like
\begin{equation}\label{hierarchy_relation_RS1}
\bar{M}_{\mathrm{Pl}}^2  = \frac{\bar{M}_5^3}{\kappa}  \;,
\end{equation}
while the RSSC relation ($C = \kappa \pi r_c$)
\cite{Giudice:05}-\cite{Kisselev:06}, \cite{Rubakov:01} is
\begin{equation}\label{hierarchy_relation_RSSC}
\bar{M}_{\mathrm{Pl}}^2  = \frac{\bar{M}_5^3}{\kappa} \, e^{2\kappa
\pi r_c} \;.
\end{equation}

The dependence of the graviton mass spectrum on $C$ can be seen
explicitly, if we rewrite eq.~\eqref{m_n} in the following
equivalent form:
\begin{equation}\label{m_n_2}
m_n = x_n \kappa \, e^{C -\kappa \pi r_c} \;.
\end{equation}
Correspondingly, one gets from \eqref{Lambda_pi} that
\begin{equation}\label{Lambda_pi_2}
\Lambda_\pi = \bar{M}_5 \left( \frac{\bar{M}_5}{\kappa}
\right)^{1/2} \, \!\!\! e^{C -\kappa \pi r_c} \;.
\end{equation}

As a result, different values of the constant $C$ leads to quite
different values of the parameters and spectra of the KK gravitons.
For instance, in the RS1 model the hierarchy relation
\eqref{hierarchy_relation_RS1} needs $\kappa \sim \bar{M}_5 \sim
\bar{M}_{\mathrm{Pl}}$ with $m_n/x_n \sim 1$ TeV, while in the RSSC
model one can take $\kappa \sim 1$ GeV, $M_5 \sim 1$ TeV, that
results in $m_n/x_n \sim 1$ GeV. Let us underline that
eq.~\eqref{hierarchy_relation_RS1} \emph{does not admit} the
parameters of the model to lie in the region $\kappa \sim 1$ GeV,
$M_5 \sim 1$ TeV. Thus, from the point of view of a 4-dimensional
observer, the models with $C=0$ and $C=\kappa\pi r_c$ are quite
different.

In the limit $\kappa \rightarrow 0$, the hierarchy relation for the
flat metric is reproduced from \eqref{hierarchy_relation},
\begin{equation}\label{flat_hierarchy_relation}
\bar{M}_{\mathrm{Pl}}^2  = \bar{M}_5^3 V_1 \;,
\end{equation}
where $V_1 = 2\pi r_c$ is the ED volume.%
\footnote{Note that $C \rightarrow 0$ in this limit, since $0
\leqslant C \leqslant |\kappa| \pi r_c$.}
Simultaneously, $\Lambda_\pi \rightarrow \bar{M}_{\mathrm{Pl}}$, and
$m_n \rightarrow n/r_c$, as one can derive from \eqref{masses_eq}.


\section{Discussions of the results}
\label{sec:3}

\begin{figure}
\begin{center}
\resizebox{10cm}{!}{\includegraphics{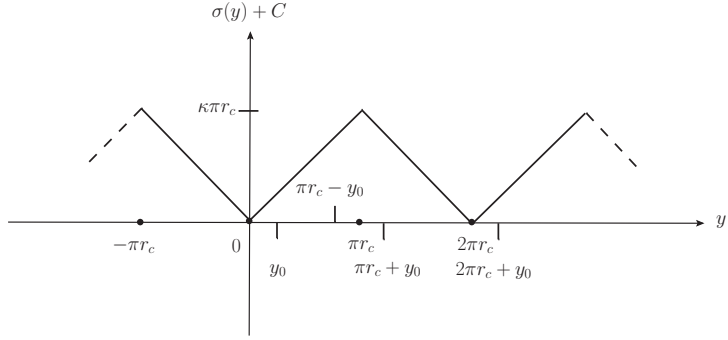}}
  \caption{Warp function $\sigma(y) + C$ given by eq.~\eqref{sigma_solution}.}
  \vspace*{-.5cm}
  \label{fig:sigma_y}
\end{center}
\end{figure}

First let us stress that the RS1 solution $\sigma_{RS}(y) = \kappa
|y|$ \eqref{sigma_RS1} can not be treated for all $y$ as
$\sigma_{RS}(y) = \kappa \, y \, \mathrm{sgn}(y)$. Namely, $|y| = y
\, \mathrm{sgn}(y)$ is assumed to be valid  in the model \emph{only}
for $|y| \leqslant \pi r_c$. Outside this region the periodicity
condition must be used \emph{before} absolute value operation
$|..|$. In other words, the value of the extra coordinate $y$
\emph{must be reduced to the interval} $[-\pi r_c, \pi r_c]$. For
instance, for $y = \pi r_c + y_0$, where $0 \leqslant y_0 \leqslant
\pi r_c$, one gets
\begin{equation}\label{absolute_value_1}
\sigma_{\mathrm{RS}}(\pi r_c + y_0) = \kappa|y_0 + \pi r_c -
\boldsymbol{2\pi r_c}| = \kappa(\pi r_c - y_0) \;.
\end{equation}
Analogously,
\begin{equation}\label{absolute_value_2}
\sigma_{\mathrm{RS}}(2\pi r_c + y_0) = |y_0 + 2\pi r_c -
\boldsymbol{2\pi r_c}| =  \kappa y_0 \;,
\end{equation}
and so on (see fig.~\ref{fig:sigma_y}).%
\footnote{Remember that $C=0$ in the RS1 model.}
%
%

The same is also true for our solution \eqref{sigma_solution}. At
first site, $\sigma(y)$ becomes a constant outside the region $0
\leqslant y \leqslant \pi r_c$. But it is not the case. Indeed,
consider, for example, $y = \pi r_c + y_0$ with $0 \leqslant y_0
\leqslant \pi r_c$. Then we have the following sequence of
equalities (for definiteness, in what follows $\kappa >0$):
\begin{align}\label{outside}
&\sigma(\pi r_c + y_0) + C = \frac{\kappa}{2} ( |y_0 + \pi r_c| -
|y_0|) + \frac{\kappa \pi r_c}{2} \nonumber \\
&= \frac{\kappa}{2} (|y_0 + \pi r_c - \boldsymbol{2\pi r_c}| -
|y_0|) + \frac{\kappa \pi r_c}{2} = \kappa(\pi r_c - y_0) \;,
\end{align}
in accordance with fig.~\ref{fig:sigma_y} and RS1 solution
\eqref{absolute_value_1}. Analogously, for $y = - y_0$ with $0
\leqslant y_0 \leqslant \pi r_c$,
\begin{align}\label{outside}
&\sigma(-y_0) + C = \frac{\kappa}{2} ( |-y_0| - |-y_0 - \pi r_c|
) + \frac{\kappa \pi r_c}{2} \nonumber \\
&= \frac{\kappa}{2} (|-y_0 | - |-y_0 - \pi r_c + \boldsymbol{2\pi
r_c}|) + \frac{\kappa\pi r_c}{2} = \kappa y_0 \;.
\end{align}

As it was already mentioned after eq.~\eqref{sigma_deriv_solution}
the functions $\varepsilon (x)$ can be treated in a standard manner
only for $0 < |x| \leqslant \pi r_c$. Outside this region the
periodicity condition should be imposed first. For instance, we
obtain for $0 < \epsilon < \pi r_c$%
\footnote{In contrast to a naive expectation
$\int_{-\epsilon}^\epsilon dy \delta (y - \pi r_c) = 0$.}
\begin{align}\label{delta_int_correct}
& \int_{-\epsilon}^\epsilon dy \delta (y - \pi r_c) = \frac{1}{2}
[\varepsilon(\epsilon - \pi r_c) - \varepsilon(-\epsilon - \pi r_c)]
\nonumber \\
&= \frac{1}{2} [\varepsilon(\epsilon - \pi r_c) -
\varepsilon(-\epsilon - \pi r_c + \boldsymbol{2\pi r_c})] = -1 \;,
\end{align}
that results in
\begin{align}\label{derivative}
\int_{-\epsilon}^\epsilon dy \sigma''(y) &= 2 \sigma'(\epsilon) =
\kappa \left[ \lambda_1 \!\! \int_{-\epsilon}^\epsilon dy \delta (y)
+ \lambda_2 \!\! \int_{-\epsilon}^\epsilon dy \delta (y - \pi r_c)
\right]
\nonumber \\
&= \kappa (\lambda_1 - \lambda_2) = 2 \kappa \;.
\end{align}
Thus, we get the correct result $\sigma'(\epsilon) = \kappa$.
Analogously, we find $\sigma'(\pi r_c - \epsilon) = \kappa
(\lambda_1 - \lambda_2) = \kappa$. The point $y=\epsilon$ ($y = \pi
r_c - \epsilon$) lies in between the branes. That is why,
$\sigma'(\epsilon)$ ($\sigma'(\pi r_c - \epsilon)$) is defined by
\emph{both} $\lambda_1$ \emph{and} $\lambda_2$. This effect is one
more manifestation of the symmetry with respect to the branes. Note
that in the RS1 model $\sigma'_{\mathrm{RS}}(\epsilon)$ is defined
by one brane only, that requires $(\Lambda_{1})_{\mathrm{RS}}$
\eqref{RS1_Lambdas_fine_tuning} to be twice as large as
$\Lambda_{1}$ \eqref{Lambdas_fine_tuning}.

Starting from eq.~\eqref{sigma_deriv_solution_gen}, one can derive a
compact expression for $\sigma'(y)$. Let $y = y_0 + (2k +1)$, where
$0 < y_0 < \pi r_c$, $k = 0, \pm 1, \ldots$. Since
\begin{align}\label{odd_pi_r}
\varepsilon (y_0 + (2k +1)\pi r_c) &= \varepsilon (y_0 + (2k +1)\pi
r_c - \boldsymbol{2(k+1)\pi r_c}) = -1 \;, \nonumber \\
\varepsilon (y_0 - \pi r_c + (2k +1)\pi r_c) &= \varepsilon (y_0 +
2k\pi r_c - \boldsymbol{2k\pi r_c}) = 1 \;,
\end{align}
we find that
\begin{equation}\label{sigma_deriv_solution_odd}
\sigma'(y_0 + (2k +1)\pi r_c) = \frac{\kappa}{2} (-\lambda_1 +
\lambda_2) = -\kappa \;.
\end{equation}
Analogously, we get for $y = y_0 + 2k\pi r_c$, $k = 0, \pm 1,
\ldots$,
\begin{equation}\label{sigma_deriv_solution_odd}
\sigma'(y_0 + 2k\pi r_c) = \frac{\kappa}{2} (\lambda_1 - \lambda_2)
= \kappa \;.
\end{equation}
Two last formulas can be combined into a compact one ($y \neq n \pi
r_c$, $n=0, \pm 1, \ldots$)
\begin{equation}\label{sigma_deriv_compact}
\sigma'(y) = \kappa \, \varepsilon (\sin (y/r_c)) \;.
\end{equation}
Equation \eqref{sigma_deriv_compact} results in relation
$\sigma'(-y) = - \sigma'(y)$.

The $Z_2$ symmetry of $\sigma(y)$ can be shown as follows:
\begin{align}\label{Z2_symmetry}
\sigma(-y) &= \frac{\kappa}{2} (|-y| - |-y - \pi r_c| ) +
\frac{\kappa\pi
r_c}{2} - C \nonumber \\
&= \frac{\kappa}{2} (|-y| - |-y - \pi r_c \boldsymbol{+ 2\pi r_c}| )
+ \frac{\kappa\pi r_c}{2} - C \nonumber \\
&= \frac{\kappa}{2} (|y| - |y - \pi r_c|) + \frac{\kappa \pi r_c}{2}
- C  = \sigma(y) \;.
\end{align}

The shift $\sigma(y) \rightarrow \sigma(y) - C$ is the change of
four-dimensional part of the metric \eqref{RS_background_metric}, namely%
\footnote{Correspondingly, four-dimensional interval changes as
$ds_4^2 \rightarrow ds_4^2 e^{2C}$.}
\begin{equation}\label{metric_change}
g_{\mu\nu} \rightarrow  g_{\mu\nu} e^{2C}.
\end{equation}
The Einstein tensor $R_{\mu\nu} - (1/2)g_{\mu\nu}R$ is invariant
under such a transformation (remember that $C$ is a constant). As
for the energy-momentum tensor, it is scale-invariant only for
massless fields. For instance, the energy-momentum tensor of the
massive scalar field,
\begin{equation}\label{e-m_tensor}
T_{\mu\nu} = \partial_\mu \phi \, \partial_\nu \phi - \frac{1}{2} \,
g_{\mu\nu} \! \left[ g^{\alpha\beta}  \partial_\alpha \! \phi \,
\partial_\beta \phi  - m^2 \phi^2 \right] ,
\end{equation}
\emph{is not} scale-invariant due to the third term in
\eqref{e-m_tensor}. In general, theories with massive fields are not
invariant under transformation \eqref{metric_change}.

Consider the effective 4-dimensional gravity action on the TeV brane
(with radion term omitted). It looks like (see, for instance,
\cite{Boos:02})
\begin{equation}\label{gravity_TeV_action}
S_{\mathrm{eff}} = \frac{1}{4} \sum_{n=0}^{\infty} \int \! d^4x \!
\left[ \partial_\mu h^{(n)}_{\varrho\alpha}(x) \partial_\nu
h^{(n)}_{\delta \lambda} (x) \, \eta^{\mu\nu} - m_n^2
h^{(n)}_{\varrho\alpha} (x) h^{(n)}_{\delta \lambda}(x) \right] \!
\eta^{\varrho\delta} \eta^{\alpha\lambda} \;.
\end{equation}
The shift $\sigma(y) \rightarrow \sigma(y) - C$ can be also regarded
as the rescaling of four-dimensional coordinates (see also
\cite{Rubakov:01})
\begin{equation}\label{x_transformation}
x^{\mu} = e^{C} x'^\mu \;,
\end{equation}
but then \emph{without change} of the metric. Let us stress that
\eqref{x_transformation} is not a particular case of general
coordinate transformation in gravity, since the metric tensor
$g_{\mu\nu}$ remains fixed.

The invariance of the action \eqref{gravity_TeV_action} under
transformation \eqref{x_transformation} needs rescaling of the
graviton fields and their mass: $h^{(n)}_{\mu\nu} = e^{-C}
h'^{(n)}_{\mu\nu}$, $m_n = e^{-C} m'_n$. We see that the theory of
massive KK gravitons is not scale-invariant. Only its zero mass
sector (standard gravity) remains unchanged.

Thus, one must conclude that warp functions $\sigma_1(y)$ and
$\sigma_2(y) = \sigma_1(y) - C$  result in two \emph{non-equivalent}
4-dimensional theories.%
\footnote{For the particular values of $C$, it was explicitly
demonstrated in the end of Section~\ref{sec:2}.}
As an illustration, the transition from the RS1 scenario to the RSSC
scenario assumes the shift $\sigma(y) \rightarrow \sigma(y) -
\pi\kappa r_c$. Correspondingly, the equation for the graviton
masses in the RS1 model,
\begin{equation}\label{masses_RS1}
m_n \simeq x_n \kappa \, e^{-\kappa\pi r_c} \;,
\end{equation}
transforms into equation in the RSSC model:
\begin{equation}\label{masses_RSSC}
m'_n \simeq m_n e^{\kappa\pi r_c} = x_n \kappa\;,
\end{equation}
in accordance with the results of
refs.~\cite{Giudice:05}-\cite{Kisselev:06},
\cite{Rubakov:01}-\cite{Boos:02}.

Recently, it was shown that an excess in the diphoton invariant mass
spectrum seen in 13 TeV data at ATLAS~\cite{diphoton_ATLAS} and
CMS~\cite{diphoton_CMS} is consistent with a warped compactification
\cite{Giddins:16}-\cite{Dillon:16}. It particular, it was supposed
\cite{Giddins:16} that the diphoton resonance at the LHC can be
produced as follows:
\begin{equation}\label{diphoton_prod}
pp \rightarrow h^{(1)} \rightarrow \gamma\gamma \;.
\end{equation}
Here $h^{(1)}$ is the lowest graviton KK mode with the mass
\begin{equation}\label{lowest_grav_mass}
m_1 = x_1 \kappa \;,
\end{equation}
where $x_1 = 3.83$ is the first zero of the Bessel function
$J_1(x)$. Thus, $\kappa = 196$ GeV for $m_1 = 750$ GeV
\cite{Giddins:16}. A comparison between \eqref{lowest_grav_mass},
\eqref{m_n_2} shows that the physical framework used in
\cite{Giddins:16} corresponds to the warped compactification
scenario with $C = \kappa \pi r_c$ (for details, see Section~2).

In ref.~\cite{Kisselev:14} the $p_\bot$ distribution for the
dielectron production at the LHC was calculated in such a scenario.
By comparing theoretical predictions with the LHC data at 7 and 8
TeV, the following lower bound on $\bar{M}_5$ was obtained%
\footnote{In \cite{Kisselev:14} the bound was presented for the
5-dimensional Planck mass $M_5 = \bar{M}_5 (2\pi)^{1/3}$.}
\begin{equation}\label{M5_bound}
\bar{M}_5 > 3.44 \mathrm{\ TeV} \;.
\end{equation}
Then for $C = \kappa \pi r_c$ we get from \eqref{Lambda_pi_2},
\eqref{M5_bound}:
\begin{equation}\label{lambda_bound}
\Lambda_\pi > 14.4 \mathrm{\ TeV} \;.
\end{equation}
This inequality is not in contradiction with (although, not close
to) the best-fit value $\Lambda_\pi \approx 60$ TeV obtained in
\cite{Giddins:16}.

It follows from eqs.~\eqref{m_n_2}, \eqref{Lambda_pi_2} that for
\emph{any} $C$
\begin{equation}\label{M5_over_kappa}
\frac{\bar{M}_5}{\kappa} = \left( \frac{\Lambda_\pi x_1}{m_1}
\right)^{2/3} .
\end{equation}
Putting $m_1 = 750$ GeV and $\Lambda_\pi = 60$ TeV, we find
\begin{equation}\label{M5_over_kappa_num}
r = \left( \frac{\kappa}{\bar{M}_5} \right)^{3/2} \! \simeq 0.003
\;.
\end{equation}
The values of the parameters $\bar{M}_5$ and $\kappa$ taken
separately depend on the particular RS-like scenario defined by the
constant $C$. Let us put
\begin{equation}\label{C_value}
C = a \kappa \pi r_c \;,
\end{equation}
where $0 < a < 1$. Then we get from \eqref{hierarchy_relation},
\eqref{m_n_2}:
\begin{equation}\label{kappa_exp}
\kappa = \bar{M}_{\mathrm{Pl}} \left( \frac{m_1}{x_1
\bar{M}_{\mathrm{Pl}}} \right)^a \!\!r^{1-a} \;.
\end{equation}
This equation results in $\kappa = 0.003 \bar{M}_{\mathrm{Pl}}$ for
the RS1 scenario ($a=0$), and  $\kappa = 196$ GeV for the RSSC
scheme ($a=1$). The gravity scale $\bar{M}_5$ is equal to $0.14
\bar{M}_{\mathrm{Pl}}$, and 8.9 TeV, respectively.


\section{Conclusions}

To summarize, we have studied the space-time with non-factorizable
geometry in four spatial dimensions with two branes (RS scenario).
It has the warp factor $\exp[\sigma(y)]$ in front of
four-dimensional metric. The generalization of the original RS
solution of the Einstein-Hilbert equations for the function
$\sigma(y)$ is obtained \eqref{sigma_solution} which: (i) obeys the
orbifold symmetry $y \rightarrow - y$; (ii) makes the jumps of
$\sigma'(y)$ on both branes; (iii) has the explicit symmetry with
respect to the branes; (iv) includes the constant $C$ ($0 \leqslant
C \leqslant |\kappa| \pi r_c$). This constant can be used for model
building within the framework of the general RS scenario.

Since our expression for $\sigma(y)$ is symmetric with respect to
the brane positions, the brane tensions appeared to be the factor of
two different than the RS1 tensions.

As a by-product, the compact analytical expression for $\sigma'(y)$
is obtained \eqref{sigma_deriv_compact}.

It is worthy to note that an explicit expression which makes the
jumps of $\sigma'(y)$ on both branes was presented in
\cite{Dominici:03},
\begin{equation}\label{sigma_two_jumps}
\sigma_{\mathrm{DGGT}}(y) = \kappa \{y [2\,\theta(y) - 1] - 2 (y -
\pi r_c)\,\theta(y - \pi r_c) \}  + \mathrm{constant} \;.
\end{equation}
However, contrary to our formula \eqref{sigma_solution}, this
expression is neither symmetric in variable $y$ nor invariant with
respect to the interchange of the branes.

Some recent results related to the interpretation of the excess in
the diphoton invariant mass spectrum at 13 TeV in terms of the
warped compactification are briefly discussed.



\section*{Acknowledgements}

The author is indebted to I.~Antoniadis, M.L. Mangano, V.A.~Petrov
and V.O.~Soloviev for fruitful discussions.




\end{document}